\pdfoutput=1
\documentclass[twocolumn]{aastex61}
\usepackage[space]{grffile}
\usepackage{amsfonts,amsmath,amssymb}
\usepackage{url}
\usepackage{latexsym}
\usepackage{hyperref} 
\usepackage{graphicx}
\usepackage{natbib}
\usepackage[utf8]{inputenc}
\usepackage{fancyref}
\usepackage{multirow}
\hypersetup{colorlinks=false,pdfborder={0 0 0},}

\begin{document}

\title{\textit{Realfast}: Real-Time, Commensal Fast Transient Surveys with the Very Large Array}
\shorttitle{\textit{realfast}}
\shortauthors{Law et al.}

\author[0000-0002-4119-9963]{C.~J.~Law}
\affiliation{Department of Astronomy and Radio Astronomy Lab, University of California, Berkeley, CA 94720, USA}

\author{G.~C.~Bower}
\affiliation{Academia Sinica Institute of Astronomy and Astrophysics, 645 N. A'ohoku Place, Hilo, HI 96720, USA}

\author{S.~Burke-Spolaor}
\affiliation{National Radio Astronomy Observatory, Socorro, NM 87801, USA}
\affiliation{Department of Physics and Astronomy, West Virginia University, Morgantown, WV 26506, USA}
\affiliation{Center for Gravitational Waves and Cosmology, West Virginia University, Chestnut Ridge Research Building, Morgantown, WV 26505}

\author{B.~J.~Butler}
\affiliation{National Radio Astronomy Observatory, Socorro, NM 87801, USA}

\author{P.~Demorest}
\affiliation{National Radio Astronomy Observatory, Socorro, NM 87801, USA}

\author {A.~Halle}
\affiliation{Department of Astronomy and Radio Astronomy Lab, University of California, Berkeley, CA 94720, USA}

\author{S.~Khudikyan}
\affiliation{Jet Propulsion Laboratory, California Institute of Technology, Pasadena, CA 91109, USA}

\author{T.~J.~W.~Lazio}
\affiliation{Jet Propulsion Laboratory, California Institute of Technology, Pasadena, CA 91109, USA}

\author{M.~Pokorny}
\affiliation{National Radio Astronomy Observatory, Socorro, NM 87801, USA}

\author{J.~Robnett}
\affiliation{National Radio Astronomy Observatory, Socorro, NM 87801, USA}

\author{M.~Rupen}
\affiliation{National Research Council of Canada, Herzberg Astronomy and Astrophysics, Dominion Radio Astrophysical Observatory, P.O. Box 248, Penticton, BC V2A 6J9, Canada}

\begin{abstract}
Radio interferometers have the ability to precisely localize and better characterize the properties of sources. This ability is having a powerful impact on the study of fast radio transients, where a few milliseconds of data is enough to pinpoint a source at cosmological distances. However, recording interferometric data at millisecond cadence produces a terabyte-per-hour data stream that strains networks, computing systems, and archives. This challenge mirrors that of other domains of science, where the science scope is limited by the computational architecture as much as the physical processes at play. Here, we present a solution to this problem in the context of radio transients: \textit{realfast}, a commensal, fast transient search system at the Jansky Very Large Array. \textit{Realfast} uses a novel architecture to distribute fast-sampled interferometric data to a 32-node, 64-GPU cluster for real-time imaging and transient detection. By detecting transients \textit{in situ}, we can trigger the recording of data for those rare, brief instants when the event occurs and reduce the recorded data volume by a factor of 1000. This makes it possible to commensally search a data stream that would otherwise be impossible to record. This system will search for millisecond transients in more than 1000 hours of data per year, potentially localizing several Fast Radio Bursts, pulsars, and other sources of impulsive radio emission. We describe the science scope for \textit{realfast}, the system design, expected outcomes, and ways real-time analysis can help in other fields of astrophysics.
\end{abstract}


\section{Introduction}
The study of fast radio transients is important for a number of fundamental problems in astrophysics, including the discovery of new classes of object like the Fast Radio Burst \citep[\hbox{FRB}, ][]{2007Sci...318..777L}, detecting the hidden baryonic matter in the intergalactic medium \citep{2014ApJ...780L..33M}, and testing general relativity via the formation and evolution of compact objects \citep{2006Sci...314...97K}. Some of these systems are very rare, so large surveys are conducted for a chance to find a single source. However, even a single discovery can produce a great scientific payoff by offering a unique view of the physics at play.

Recent technological advances, exemplified by the development of the Jansky Very Large Array (VLA), have opened access to fast transients through millisecond timescale radio interferometric imaging. While large, single-dish telescopes have pioneered the field, interferometers will transform it through their ability to precisely localize, efficiently survey, and better characterize source properties. For example, the 100-meter Green Bank Telescope and VLA have comparable sensitivities, but the VLA surveys 16 times more sky while localizing sources 10 to 300 times more precisely. In this way, images from an interferometer function as a high-resolution, multi-beam receiver. However, this powerful capability can only be exploited if data rates larger than 1 TB hour$^{-1}$\ can be searched for hundreds to thousands of hours. A new paradigm is needed to solve the ``needle in the haystack'' problem with such massive data streams.

\textit{Realfast} will solve this problem by integrating a compute cluster with the VLA for \textit{in situ} analysis in support real-time decision making. \textit{Realfast} is a 32-node, 64-GPU compute cluster that will search images generated on timescales from 1~ms to 1~minute. \textit{In situ} analysis allows triggered recording of a parallel data stream for later analysis and the rapid announcement of candidates for follow-up observing. By integrating with a high-speed, duplicate data stream of the VLA, this system will turn each observation into a fast transient survey, ultimately encompassing thousands of hours per year. 

In Section \ref{sec:transients}, we describe the science potential of fast radio transients and the challenge in finding them. Section \ref{sec:system} presents the \textit{realfast} transient search algorithm and system design. In Section \ref{sec:status}, we discuss the development status and expected performance.

\section{The Promise and Challenge of Fast Radio Transients}
\label{sec:transients}

\subsection{Fast Radio Transient Science}
\label{sec:science}

Radio transients are commonly considered ``fast'' if they last less than about 1 second. This time scale crudely separates slowly evolving synchrotron transients \citep{1994ApJ...426...51R, 2015MNRAS.446.3687P} from fast, coherent emission processes. The best known class of fast radio transient is the pulsar, a rotating neutron star that is detected through brief intense pulses of light \citep{1968Natur.217..709H, 1975ApJ...195L..51H}. On fast timescales, radio light propagation effects, such as dispersion\footnote{The cold plasma dispersion law is characterized by a delay of $\tau=4.15~\rm{ms}~\left(\rm{DM}/\nu^2\right)$, where DM refers to the dispersion measure in units of pc cm$^{-3}$\ and $\nu$\ is the observing frequency in units of GHz.} can be measured. Propagation effects have been used to infer the distances and sizes of transients \citep{2014ApJ...780L...2B} and model Galactic structure \citep{2002astro.ph..7156C}.

The newest class of fast radio transient is the FRB, a highly-dispersed, millisecond transient. Roughly two dozen FRBs have now been detected by radio telescopes around the world\footnote{See the FRBCat at \url{http://www.astronomy.swin.edu.au/pulsar/frbcat} \citep{2016PASA...33...45P}.}. The radio signal is similar to that of a pulsar (impulsive, dispersed, scattered), so similar algorithms have been used to detect and study both classes of object. We recently used a prototype version of \textit{realfast} at the VLA to precisely localize an FRB and identify its host galaxy at a redshift of 0.193 \citep{LOC, OPT}. This has shown that FRBs function like pulsars, but probe the intergalactic and circumgalactic media \citep[IGM, CGM;][]{2013ApJ...776..125M, 2016ApJ...824..105A,2014ApJ...780L..33M}. With many FRB localizations at high redshift, they could even be used to constrain the dark energy equation of state \citep{2014PhRvD..89j7303Z}.

A similar approach can be applied in the study of Galactic radio transients. The first demonstration a fast radio imaging search algorithm was made in a study of a “rotating radio transient” (RRAT), a potentially new class of pulsar \citep{2006Natur.439..817M, 2006ApJ...645L.149W, 2009MNRAS.400.1439L}. We used fast imaging at the VLA to make the first interferometric localization of a RRAT that excluded its association with optical/IR counterparts expected under the magnetar model \citep{2012ApJ...760..124L}. Neutron stars in the Galactic center, globular clusters, or nearby galaxies \citep{2010ApJ...715..939M, 2013MNRAS.428.2857R} may be detectable as millisecond transients. By localizing them to arcsecond precision, we can associate them with multiwavelength counterparts to infer a distance, find companions, or measure properties of their local environment.

\textit{Realfast} can also be applied to the search for periodic radio emission. Single-dish telescopes currently do both pulsar discovery via by searching for periodic emission and characterization by modeling pulse arrival times. Timing analysis provides precide astrometry, measures binary motion, and determines intrinsic spin-down rate of the pulsar.  However, regular observations over at least one year are required to separate astrometric effects from spin-down and long-period binary motion \citep{2001PhDT.......123R, 2016ApJ...828....8D}.  Modern pulsar surveys are discovering pulsars so rapidly that many can not be completely followed-up with timing observations, potentially missing exotic systems with great potential to test general relativity \citep[e.g., neutron star--black hole binary;][]{2006PhRvD..73f3003R, 2011MNRAS.415.3951F}. An arcsecond localization that comes via interferometric detection instantly limits the range of spin-down and binary properties by constraining the astrometric timing parameters.  An arcsecond localization corresponds to timing precisions on the order of milliseconds ($\delta t \approx 1~ \rm{AU} \times \delta\theta$), which makes the VLA ideal for sorting ordinary pulsars from more exotic systems.


\subsection{The Interferometric Search Problem}
\label{sec:problem}

It is clear that arcsecond localization can dramatically expand the science scope of fast radio transients. Large, single-dish telescopes localize with a precision of several arcminutes, which is too poor to uniquely identify optical counterparts \citep{2017arXiv170502998E}. Interferometers further improve on single-dish data by providing measurements of flux and spectra unaffected by uncertainties about source position within the telescope beam, as well as robust rejection of bad data. 

\begin{figure}[ht]
\begin{center}
\includegraphics[width=\columnwidth]{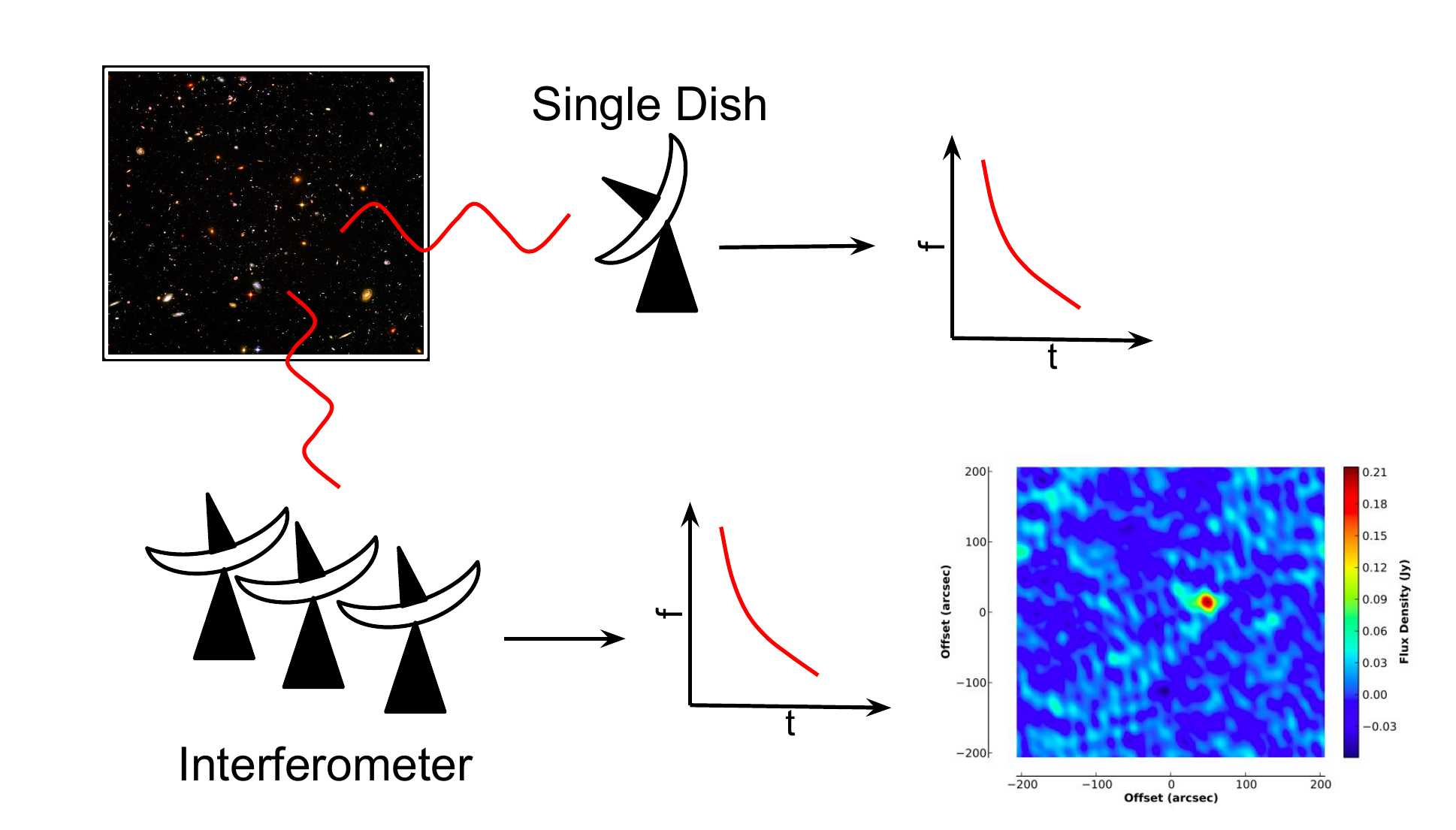}
\caption{A schematic representation of interferometric and single-dish approaches to transient detection. The single-dish approach produces a single data stream (or small number of data streams, for multi-beam receivers) with dimensions of time versus frequency that is searched for impulsive signals. An interferometer can produce images for each integration that is searched for a source. The image shows our first demonstration of this concept, which blindly detected and localized RRAT J0628+0909 \citep{2012ApJ...760..124L}. Since then, we have used this concept to blindly detect and localize FRB 121102 \citep{LOC, 2017arXiv170507553L}. \label{fig:int}}
\end{center}
\end{figure}

Using interferometers to study fast transients presents a significant data analysis challenge. Millisecond sampling of interferometric data produces data at rates higher than 1 TB hour$^{-1}$, roughly 100 times larger than is typical \citep{2015ApJ...807...16L}. The transient search requires searching each integration with hundreds of matched filters to capture impulses over a range of dispersions and pulse widths (see Figure \ref{fig:int}). The traditional collect-and-analyze observing model is not tenable for this science case.

Fast transient science is growing just as interferometers are coming to dominate radio astronomy \citep{2009IEEEP..97.1507D, 2009arXiv0910.2935B, 2014SPIE.9145E..22B}. An approach taken by many newly-designed interferometers is beamforming, in which the interferometric signal is reduced to a number of data streams corresponding to different sky positions within the interferometer's field of view. Each data stream has dimensions of time versus frequency, thus can be processed independently as if it were the output of a single-dish instrument \citep{1996PASA...13..243S}. The resulting data rate, and computational cost, of processing the entire field of view thus scales with the total number of pixels in the image, $N_{\rm{pix}}\propto (l/D)^2$, where $l$\ is the length of the longest interferometer baseline and $D$\ is the diameter of each dish. An alternate approach, more traditionally used for interferometric imaging, is to compute ``visibilities,'' the correlation of electromagnetic wave between pairs of antennas. The visibility data stream encodes the same physical information as beamformed data, but is produced at a rate that scales only as the number of baselines (antenna pairs), $N_{\rm{bl}}=N_{\rm{ant}}(N_{\rm{ant}}-1)/2$, independent of $N_{\rm{pix}}$. For sparse interferometers like the VLA, tens of thousands to tens of millions of beams are required to cover the full field of view, so the visibility data rate can be several orders of magnitude smaller than the corresponding beamformed data rate. By performing processing operations such as dedispersion in the visibility domain there is a proportional savings in computational cost. However, closely-packed interferometers such as UTMOST \citep{2016MNRAS.458..718C} or CHIME \citep{2017arXiv170204728N} may prefer a beamforming approach.

\textit{Realfast} addresses the challenges of fast interferometric data analysis in three ways:
\begin{enumerate}
 \item \textit{In situ} analysis --- \textit{Realfast} integrates the transient search with the data recording system for real-time data triage. This will reduce the recorded data rate by a factor of $10^3$ and allow rapid response to new transient detections.
 \item Commensal observing --- \textit{In situ} analysis also makes it possible to distribute a high-speed copy of all VLA observations to the transient search system.
 \item Visibility-domain processing and FFT imaging --- The computational cost of this approach is more efficient than beamforming for interferometers with high spatial resolution like the VLA.
\end{enumerate}
Furthermore, \textit{realfast} software and data will be open, making this novel capability at the world's most sensitive radio interferometer available to astronomers at large.

\section{System Description}
\label{sec:system}

\subsection{Algorithm}

Since radio interferometers fully sample electromagnetic radiation, a wide variety of techniques exist to search for transients \citep{2011ApJS..196...16B, 2012ApJ...749..143L, 2013ApJS..206....2B, 2015A&A...579A..69O}. The principal challenges are driven by data rate, since data must be generated and searched on millisecond timescales for every pair of antennas in the array. Our approach has focused on interferometric imaging because it maintains the ideal sensitivity, is computationally efficient, and can easily be adapted to existing telescopes \citep{2011ApJ...742...12L, 2012ApJ...760..124L}.

\begin{figure*}[ht]
\begin{center}
\includegraphics[angle=270, width=2\columnwidth]{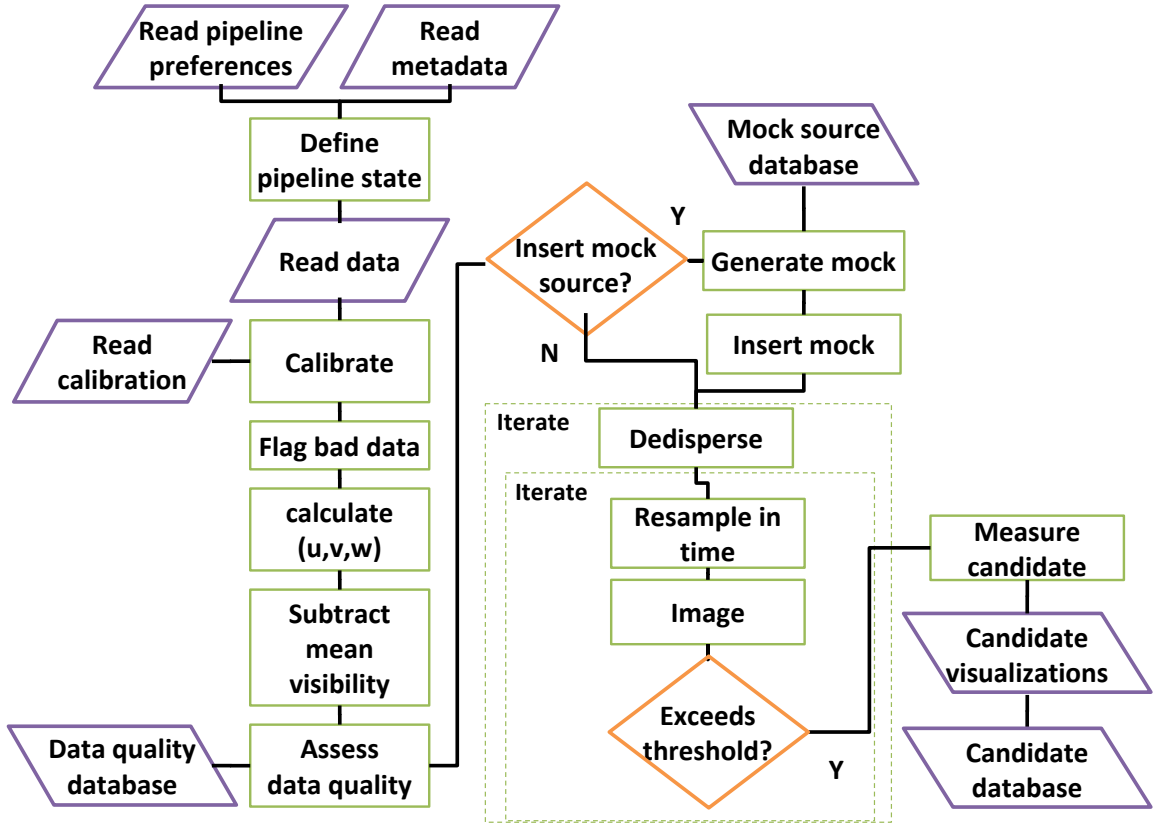}
\caption{Transient search pipeline for \textit{realfast}, known as \textit{rfpipe}. Purple, right slanted boxes denote sources of data and left slanted boxes denote output data products. Green boxes denote actions and the orange diamond shows a decision point. The dashed boxes show iteration loops over trial values of dispersion and sampling time. Other search algorithms, such for periodic sources, can be seen as a modification of the innermost iteration loop. Candidate visualizations are generated for candidates above a second, higher threshold.
\label{fig:pipe}}
\end{center}
\end{figure*}

Figure \ref{fig:pipe} shows the data flow through the search pipeline, as implemented in the Python package \textit{rfpipe}\footnote{\textit{Rfpipe} is open source software and built on other open source packages, including pwkit \citep{2017ascl.soft04001W} and astropy \citep{2013A&A...558A..33A}. All \textit{realfast} software can be found under the github organization \url{https://github.com/realfastvla}.} \citep{2017ascl.soft10002L}. The pipeline breaks the problem of studying transient sources into two pieces: detect then analyze \cite[see also][]{2015ApJ...807...16L}. By using hierarchical decision making, the pipeline dramatically reduces the data rate by saving information for a subset of the observation or only saving summary statistics.

Unlike traditional interferometric imaging \citep{2007ASPC..376..127M, 2012ascl.soft09010N}, millisecond-long integrations have modest sensitivity ($\sim$10~mJy in $\sim10$~ms). At this sensitivity, there are roughly 2 persistent sources detectable per square degree \citep{1995ApJ...450..559B} and computationally demanding imaging processes \citep{1980A&A....89..377C, 1984iimp.conf..333S} are not needed. The algorithm is further simplified by subtracting the mean visibility in time on timescales less the VLA fringe rate ($\sim$1~s). This removes constant sources, eliminates the need for source catalogs, and makes it possible to detect candidate transients by thresholding each image.

The bulk of the work in the search pipeline is in forming a matched filter for the transient signal in time and frequency. A Stokes I image is formed for each DM and time width and thresholded to search for point sources\footnote{\textit{Realfast} will also search for periodic signals by reading a time segment, transforming it to frequencies with a Fourier transform, and imaging each frequency to search for periodic signals.}. Since constant source have been subtracted, all remaining point sources must be changing on timescales less than the fringe rate of a few seconds. Point sources with S/N greater than a threshold (typically 6$\sigma$) are saved to the candidate database and those with S/N greater than a higher threshold (typically 7$\sigma$) have candidate visualizations made for follow-up inspection.

\subsection{Hardware and Software}

\begin{figure*}[ht]
\begin{center}
\includegraphics[width=2\columnwidth]{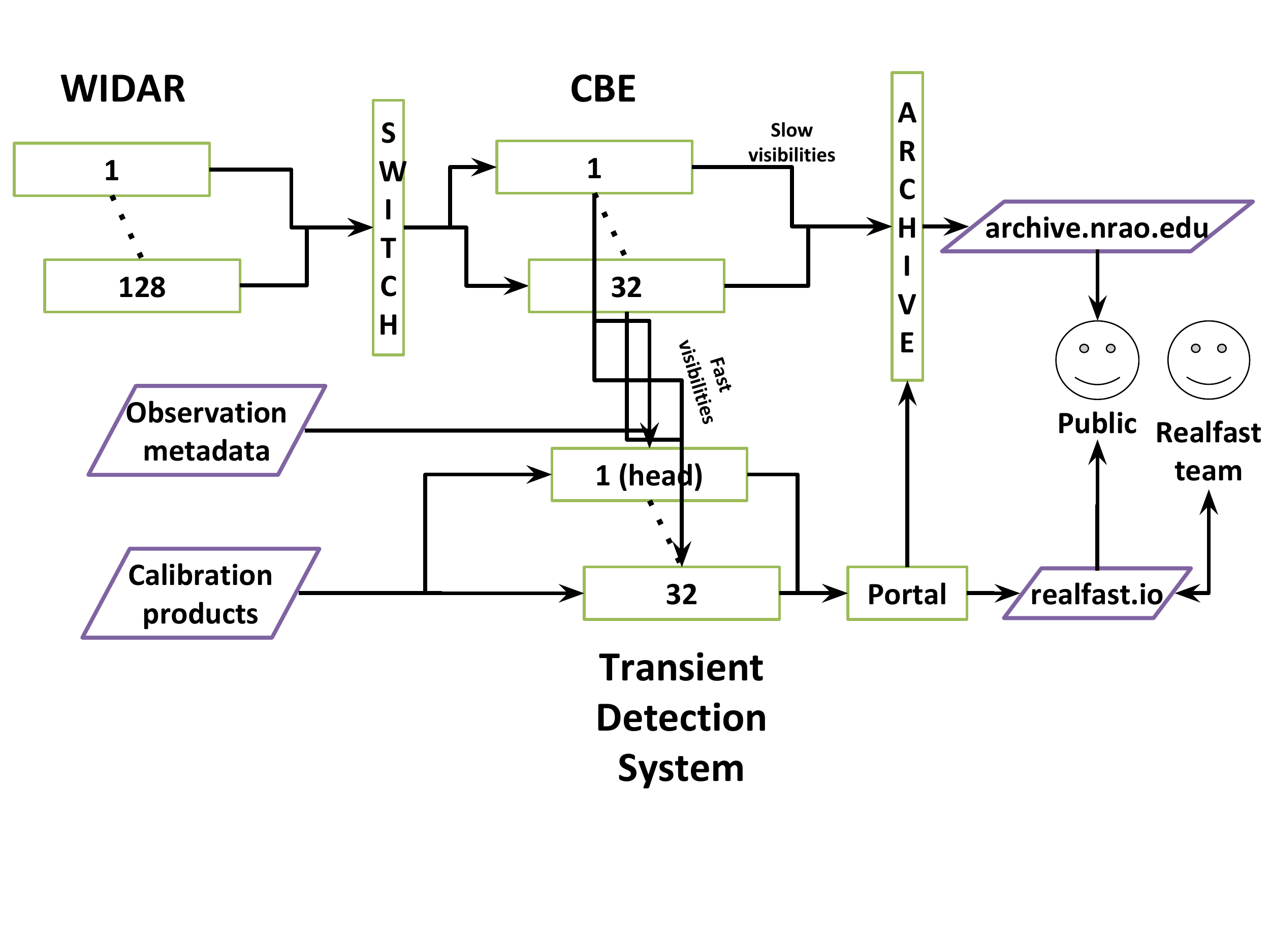}
\caption{System diagram for VLA systems, including \textit{realfast}. The VLA correlator includes WIDAR and the CBE, which together produce the raw data for both primary and commensal observers. The Transient Detection system runs the \textit{realfast} search pipeline and is managed by a process that runs on the first node of the cluster. Output products and databases are managed by the portal, which also allows team team members to access and classify new data prior to archiving. Public access to raw data is managed by the NRAO archive, but smaller data products will be served by the \textit{realfast} portal.
\label{fig:sys}}
\end{center}
\end{figure*}

A key feature of \textit{realfast} is the integration of dedicated computing with the VLA system. The VLA generates data with a large, dedicated computing system called a correlator (see Figure \ref{fig:sys}). The first stage of correlation is done in Field Programmable Gate Arrays (FPGAs) with a system called WIDAR, while the second stage is done in a commodity compute cluster known as the correlator backend \citep[CBE; ][]{2011ApJ...739L...1P}. The fast transient search will be supported by the Transient Detection (TD) system, a 32-node, 64-GPU compute cluster on the same infiniband network.

Figure \ref{fig:sys} shows how the data stream is duplicated for commensal processing. The standard data path typically gets averaged to a value defined by the primary user ($\sim$1~s). This time averaging is done in two stages, first by WIDAR and then by the CBE. \textit{Realfast} 
will configure the correlator to perform less time averaging in WIDAR, which will preserve millisecond-resolution visibilities in the CBE. Since this configuration produces more visibility data in the CBE and more are distributed over the network, both computational and network limits are important. A \textit{realfast} goal is to generate 1~ms visibilities for the entire array to search data at a rate of 1.4~GB s$^{-1}$ (see Table \ref{tab:setup} for example observing configurations). 

With fast-sampled visibility data available in the CBE, it is possible to export it over the infiniband network at rates up to 40 Gb s$^{-1}$. The distribution of visibilities is done by a new protocol for distributing and consuming VLA data via infiniband called \textit{vys} and \textit{vysmaw} \citep{2017ascl.soft10001P}. \textit{Vys} is a lightweight messaging protocol that broadcasts the availability of data in CBE memory while \textit{vysmaw} can optionally read via remote memory access protocols of infiniband networks.

\textit{Realfast} will use \textit{vys} to read data in short time segments. The segments will be the size of the inverse fringe rate (of order seconds), which is the largest size for which mean visibility subtraction is accurate. Each node will receive a segment in round robin fashion and process in slower than real time. Real-time processing is maintained over all nodes, but with a latency of (32 nodes~$\times~\sim~\rm{seconds~node}^{-1})~\approx~\sim\rm{minutes}$. This latency can be reduced by adding processing breadth (more nodes or GPUs), since each segment is independent. Segments will partially overlap in time to keep sensitivity to dispersive delays of all scales at all times. To keep the overlap to less than 10\% of the total data volume, the minimum data segment size should be at least 10 times the largest dispersive sweep of roughly 2 seconds (e.g., for DM of 2000 pc cm$^{-3}$, bandwidth of 256 MHz at frequency of 1.4 GHz).

The VLA observing system uses TCP to broadcast XML documents that define the configuration of an observation (e.g., antenna positions, correlator configuration, start time, etc.). A single node of the Transient Detection cluster will run an asynchronous process written in Python to capture these documents, parse them, and determine if they should be searched for transients. If so, then the pipeline state is defined from a set of preferences and observation metadata, as shown in Figure \ref{fig:pipe}. The state is uniquely defined for a set of metadata and preferences, such that metadata and preferences can be associated with any given candidate to reliably reproduce it offline.

Each segment of data is scheduled for processing by the \textit{distributed} library\footnote{See \url{https://github.com/dask/distributed}.}, a part of the \textit{dask} parallel computing framework. \textit{Distributed} builds a directed acyclic graph for the pipeline to find the most efficient way to schedule the individual stages of the pipeline, given constraints on memory, processors, and GPU availability. When nodes are added or removed from the cluster, \textit{distributed} automatically redistributes pipeline jobs. Many functions (e.g., dedispersion) are written as kernels with numba\footnote{See \url{http://numba.pydata.org}.} such that they can be run on multi-core CPUs. The most computationally demanding stage of processing is the 2d FFT for imaging, which is implemented on the GPU using \textit{pycuda}.

\subsection{Products and Services}
\label{sec:services}

Figures \ref{fig:pipe} and \ref{fig:sys} summarize the products and services associated with \textit{realfast}. The data products are:
\begin{enumerate}
 \item Candidate statistics database -- Statistical properties will be measured from the image and spectrum of each candidate above a threshold. These properties will summarize the properties of the candidate and in many cases help distinguish good astrophysical events from interference. Since these properties are easy to calculate and save, they are calculated at a relatively low threshold ($\sim6\sigma$) and kept indefinitely.
 \item Candidate plot -- Initially, a candidate can only be verified by visual inspection of its image and spectrum. These will be calculated for candidates above a somewhat higher threshold ($\sim7\sigma$), as they are larger and require more time to generate. They will be associated with candidates in the statistics database and kept indefinitely.
 \item Raw data cutout -- Each candidate with a plot will also be associated with raw visibility data for a small time window around the candidate. These are relatively large (several GB per candidate), so they will be deleted if they are clearly false positives. If they are worth keeping, they are archived in the standard NRAO archived and associated with a candidate in the statistics database.
 \item Data quality database -- Since \textit{realfast} intends to throw away roughly 99.9\% of the fast visibility data, we must keep a reliable measure of the amount of data searched and its quality. The data quality database will include regular measures of the fraction of configuration of the data, the amount of data flagged due to interference, and the sensitivity of a typical image.
 \item Mock source database -- The search pipeline will randomly insert mock transient sources into data as a test of its efficiency and sensitivity. The mock source is inserted after calibration, so it only tests the algorithm. This also makes it impossible to accidentally identify a mock transient as an astrophysical transient.
\end{enumerate}

Of these products, the most important is the raw visibility data that will be created for each candidate transient. We require a full reanalysis to validate any candidate transient, so raw data, observing metadata, and calibration files must be saved for offline analysis. For a typical commensal data set, we expect a raw data rate of 1.4 GB s$^{-1}$ and a size of several GB per candidate (see Figure \ref{tab:setup}). An example of this data product is available on \dataset[dataverse]{https://doi.org/10.7910/DVN/ZKESD4}, which shows nine visibility data files and associated calibration products for bursts detected from FRB 121102.

\begin{figure}[ht]
\begin{center}
\includegraphics[angle=270, width=\columnwidth]{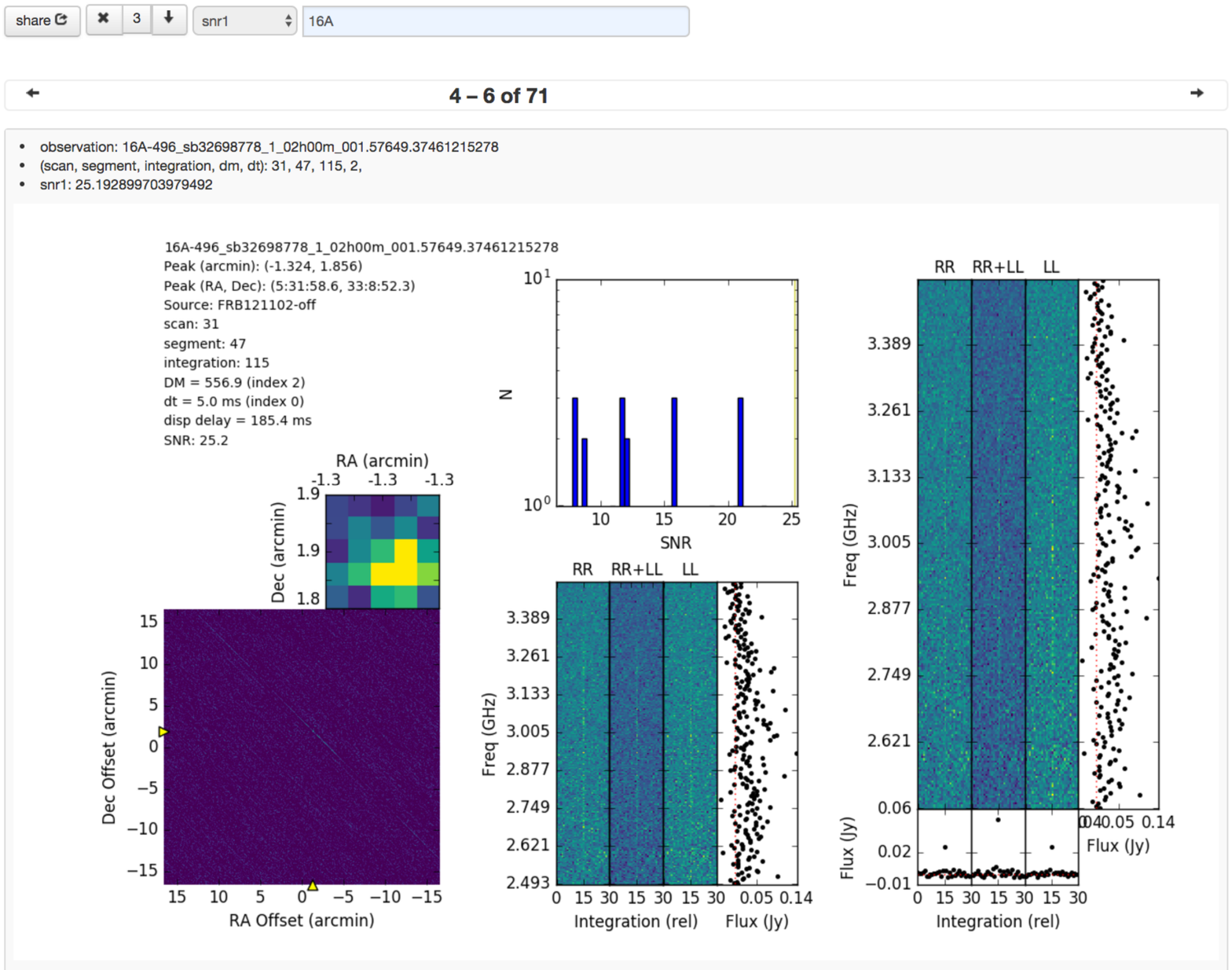}

\includegraphics[width=\columnwidth]{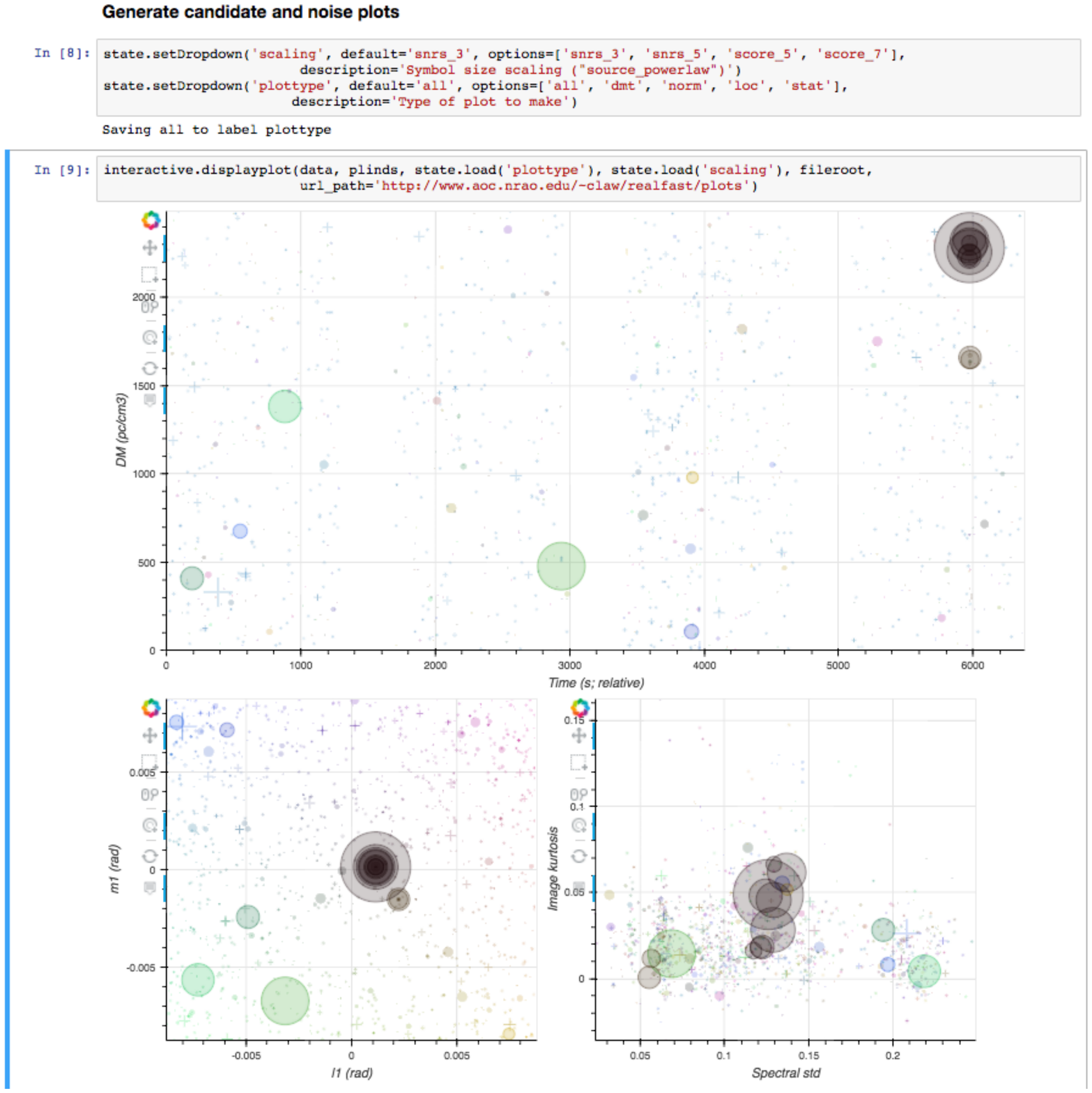}
\caption{(Top) A view of a searchable interface to the candidate database. The candidate plot shows the image and spectrum of the candidate, in this case a pulse from FRB 121102. (Bottom) A portion of a Jupyter notebook that summarizes candidate transients during an hour long observation on MJD 57970. The notebook includes interactive visualizations and can support reanalysis on demand. A static version of this notebook is available as a supplement to this paper as ``example\_notebook.ipynb''.
\label{fig:services}}
\end{center}
\end{figure}

The \textit{realfast} services are summarized in Figure \ref{fig:sys} and example views are shown in Figure \ref{fig:services}. They include: 
\begin{enumerate}
 \item Candidate search and review -- We expect to generate candidate visualizations for a few tens of events per day, most of which will be false positives triggered by terrestrial interference and thermal noise. Raw data with good candidate events must be sent to the NRAO archive, while raw data from bad candidates must be deleted. Figure \ref{fig:services} shows a web application will allow \textit{realfast} team members to search and tag candidate transients to trigger this data management process. This is supported by a database indexed with \textit{elasticsearch}\footnote{See \url{https://www.elastic.co}.}.
 \item Candidate metaanalysis via Jupyterhub -- The typical VLA observation lasts about an hour and is composed of many separate minute-scale segments with identical configuration. Figure \ref{fig:services} shows a Jupyter notebook we use to summarize candidate transients from a whole observation to provide context on overall data quality of observation. We will use Jupyterhub \citep[an extension of IPython;][]{PER-GRA:2007} to serve these and other computational notebooks to the collaboration.
 \item VOEvent server -- Astrophysical transients will be of immediate interest to the community and will be distributed by the FRB VOEvent protocol (Petroff et al, in prep; see \url{https://github.com/ebpetroff/FRB_VOEvent}).
\end{enumerate}

Of these services, the most important is the candidate review service. This service was inspired by the V-FASTR project that searched for fast radio transients in VLBA data \citep{2011ApJ...735...97W, 2016ApJ...826..223B}. Initially, our team will manually inspect each candidate to remove false positives that do not need to be archived or analyzed further, but we ultimately aim to automate the process with classification based on candidate statisics. With experience, we will reduce the archived data rate by a factor of 1000 and speed the time from discovery to confirmation.

\section{Latest Developments}
\label{sec:status}

\subsection{Status}

As of late 2017, most major \textit{realfast} software components have been either developed or prototyped. The \textit{vys} protocol has been tested in commensal observations with integration times of 50 ms. This data can be distributed to a dedicated compute node that runs the transient search pipeline running on a prototype \textit{realfast} server with two GPUs (nVidia GeForce GTX 1080) and an early version of a GPU-accelerated search pipeline. This pipeline detects bright pulsars and generates visualizations of candidate transients.

Much of the correlator and search pipeline has been prototyped in non-commensal mode over the last few years. Table \ref{tab:setup} shows example \textit{realfast} correlator configurations, including some that have been demonstrated in non-commensal (dedicated) observations for FRBs \citep{2015ApJ...807...16L, LOC}. Data was searched with an earlier version of the pipeline \emph{rtpipe} \citep{2017ascl.soft06002L}, which is designed for multi-core CPUs by dedicating each core to a set of integrations in shared memory. Profiling of the old pipeline shows that the dedispersion and imaging stages combined take about 1~ms per $512\times~512$~image. This is dominated by the FFT and scales as expected for larger images, while smaller image processing time is dominated by dedispersion. Processing time scales sublinearly with the number of cores, suggesting that memory movement reduces efficiency by as much as 50\% for 20-core CPUs.  The new pipeline will improve on this by using GPUs, a different job scheduling scheme, and a different CPU core parallelization scheme.

\begin{deluxetable*}{ccccccc}
\tablecaption{Example Observing Configurations \label{tab:setup}}
\startdata
\tablehead{
\colhead{Freq.\ Band} & \colhead{Freq.\ Center/Width} & \colhead{Channels} & \colhead{Time Resolution} & \colhead{Data rate} & \colhead{Sensitivity} & \colhead{On-sky Detection} \\
& \colhead{(GHz/MHz)} & \colhead{(ms)} & & \colhead{(MB s$^{-1}$)} & \colhead{($10\sigma$; Jy ms)} & \colhead{Rate (hours\,burst$^{-1}$)}}
L\tablenotemark{a} & 1.4/256   & 256    & 5  & 290  & 0.5  & 500--1070 \\
L                  & 1.4/256   & 256    & 1  & 1400 & 1   & 380--630 \\
S\tablenotemark{a} & 3.0/1024  & 256    & 5  & 290  & 0.2  & 190--2050 \\
S (VLASS)          & 3.0/1500  & 1024   & 50 & 120  & 0.5 & 460--5340 \\
C\tablenotemark{a} & 6.0/2048  & 256    & 5  & 290  & 0.13 & 230--10100 \\
X                  & 10.0/4096 & 256    & 5  & 290  & 0.09 & 320--38450 \\
\enddata
\tablecomments{All configurations assume 27 antennas (351 baselines) and two polarizations.}
\tablenotetext{a}{This mode commissioned in dedicated observations.}
\end{deluxetable*}

Future development will focus on testing faster commensal data streams, integrating with the VLA observing system and archiving, and improving data services in response to test observations. Our goal is to perform end-to-end commensal science observing by the end of 2017 on CPU hardware currently in place at the VLA. The dedicated GPU cluster will be purchased and installed in the first quarter of 2018.

\subsection{Performance Goals}

As a commensal project, the the \textit{realfast} science potential depends on the distribution of VLA observing time. In a typical year, the VLA observes for about 6000 hours at frequencies from 1 to 50~GHz. The antenna configuration is changed between four possible states from ``A'' (36 km baselines) to ``D'' (1 km baselines). Roughly 40\% of observations are made at frequencies less than 10~GHz, which are most sensitive to fast, coherent emission processes and have a relatively large field-of-view. About 70\% of observations make relatively light use of the correlator (continuum-like mode), which are the easiest for \textit{realfast} to use. Considering both effects, we expect a typical year of commensal observing to search $\sim$1500~hours summed over all antenna configurations.

Table \ref{tab:setup} shows example correlator configurations that we intend to search commensally, their expected sensitivity, and FRB detection rate. These rates are calculated based on the formulation of \citet{2014ApJ...792...19B} to extrapolate from one observing set up with a given FRB rate to a rate under a new observing setup. As a baseline FRB rate, we use that of \citet{2016MNRAS.460L..30C} and scale to the \textit{realfast} set-ups listed in Table \ref{tab:setup}. We make two alterations in the \citet{2014ApJ...792...19B} formulation, which is that (1) we don't take into account Galactic effects (i.e. we only consider rate-scaling effects caused by instrumentation sensitivity and channelization differences) and (2) we use a corrected version of their equation 3. As shown in Chawla et al (2017), the scaling of the relative detection rate should instead scale with frequency as:
\begin{equation}
\frac{N_1}{N_2} \propto \bigg (\frac{\nu_1}{\nu_2}\bigg)^{\alpha\gamma}
\end{equation}
\noindent where $\alpha$\ is the spectral index and $\gamma$\ is the powerlaw slope of the cumulative flux distribution.

The FRB detection rate ranges reflect uncertainty in FRB properties, such as average DM (assumed between $400-800$~pc cm$^{-3}$), intrinsic width, and scattering timescale of FRBs (between 0--3~ms at 1~GHz), a source count index of --1 \citep{2016ApJ...830...75V, 2017AJ....154..117L}, and a spectral index range of -1.0 to 1.0. The two 1.4~GHz modes show the currently-commissioned mode with 5~ms integration time and a goal of 1~ms integration time. Using the formulation as described above, we make predictions gridding across a range of combinations of these values and report the minimum and maximum outcome for detection rates.

Modes will be commissioned prior to science observing to ensure that the correlator can safely generate both primary and commensal data. The highest priority modes for \textit{realfast} will be at low frequencies (for sensitivity to coherent transients), use continuum-like spectral configurations (to minimize demand on the correlator), and compact antenna configurations (to minimize demand on the search pipeline). \textit{Realfast} will specifically target commensal searches during VLASS, an ongoing, 5500~hour sky survey from 2--4 GHz \citep{VLASS}. Based on past correlator commissioning, we conservatively estimate commensal observations can be made with 50 ms integrations. This would give a $1\sigma$~ sensitivity of 1 mJy per integration, equivalent to a $10\sigma$~ fluence limit of 0.05 Jy~ms. The VLASS detection rate translates to between 1 and 12 FRBs discovered for the entire 5500-hour survey.


\section{Conclusions}
\label{sec:con}

\textit{Realfast} is a system that will commensally search VLA data for fast radio transients. We expect to search thousands of hours of data from the world's most sensitive radio interferometer, potentially localizing several FRBs to arcsecond precision and associating them with host galaxies. This potential has already been demonstrated in the near-real-time detection and precision localization of an FRB by a prototype version of \textit{realfast}. The complete system will open discovery potential by continually searching for fast transients toward VLA targets such as galaxies, YSOs, AGN, TDEs, SNe, and GRBs. \textit{Realfast} results and data products will be shared in near real time to bring multiplicative benefits to VLA observations.

This system is an example of how \textit{in situ} analysis can open access data intensive science. By integrating dedicated computing with the observing system, \textit{realfast} can triage the data flow through a series of decisions that can be made based on available computation and a well-defined model of the transient signal. This is supported by a flexible and extensible design that allows other kinds of commensal analysis, such as periodicity searches for pulsars or flagging to improve the data quality for primary observers. As one of the first large commensal data analysis concepts deployed at an interferometer, \textit{realfast} can drive design of new interferometers under development around the world.

The growth of astronomical surveys has produced a torrent of data that has driven discovery via machine learning \citep{2012PASP..124.1175B} and inference \citep{2010AJ....139.1782L, 2015ApJ...806..215F}. In contrast, \textit{realfast} demonstrates the strategy of asking specific questions of data prior to recording it. This strategy is not new to astronomy, as it was pioneered in data intensive fields such as astro-particle physics \citep{2004PhRvL..93d1101G, 2005Natur.435..313F}. However, as instruments for time-domain astrophysics continue to grow in sophistication \citep{2009arXiv0912.0201L, 2016MNRAS.457.3036H}, triage will continue to be an effective strategy to manage the data deluge.

\bibliographystyle{apj}

\section*{Acknowledgements}
We thank Bridget Andersen, the NRAO, and especially the VLA staff for their support of \textit{realfast} development. 

The National Radio Astronomy Observatory is a facility of the National Science Foundation operated under cooperative agreement by Associated Universities, Inc..
Part of this research was carried out at the Jet Propulsion Laboratory, California Institute of Technology, under a contract with the National Aeronautics and Space Administration.
\textit{Realfast} is supported by the NSF Advanced Technology and Instrumentation program under award 1611606. 

\facility{EVLA}

\software{vysmaw \citep{2017ascl.soft10001P}, rfpipe \citep{2017ascl.soft10002L}, pwkit \citep{2017ascl.soft04001W}, astropy \citep{2013A&A...558A..33A}}

\bibliography{fasttrants.bib}

\end{document}